\documentstyle[aps,multicol,epsf,graphicx]{revtex}
\makeatletter
\def\thepage{\@arabic\c@page}
\def\@pnumwidth{2em}

\makeatother

\begin{document}
\bibliographystyle{unsrt}


\newlength{\oldcolwidth}

\newlength{\newwidth}
\setlength{\newwidth}{85mm}

\newcommand{\hostgueststructure}{
\setlength{\oldcolwidth}{\columnwidth}
\setlength{\columnwidth}{\newwidth}
\hspace{5mm}\begin{figure}
\begin{center}
\includegraphics[scale=0.55,angle=-90]{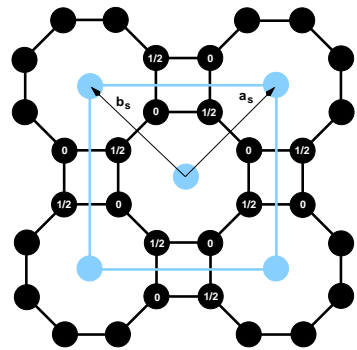}
\end{center}
\caption{\emph{Schematic representation of the Ba IV structure. The host
  atoms (dark symbols) with chains of guest atoms (light symbols) are
  projected on to the $ab$-plane. The host atoms are labeled with
  their $z$ coordinates. The arrows labelled $a_s$ and $b_s$ are
  projections of the supercell vectors used in the calculations. 
  At 12.1GPa the host cell has dimensions 
 \host{a}=\host{b}=8.42\AA{} and \host{c}=4.74\AA. 
Full description of the observed guest atoms involves three unit
cells. One is partially disordered and
tetragonal with the same $a$ and $b$ as the host structure but with
\guestc{c}=3.41\AA.   On ordering, the guest structure 
can undergo a monoclinic distortion to one of four similar cells,
related by 90$^\circ$ rotations about the tetragonal axis,  with
\guestm{a}=8.46\AA, \guestm{b}=\host{a}, \guestm{c}=3.43\AA{}  and
\guestm{\beta}=96.15$^\circ$. This corresponds to displacement
of atoms in adjacent chains by 0.47\AA{}  along $c$.  
}} 
\label{fig:hostguest}
\end{figure}
\setlength{\columnwidth}{\oldcolwidth}

}

\newcommand{\energygraph}{
\setlength{\oldcolwidth}{\columnwidth}
\setlength{\columnwidth}{\newwidth}
\begin{figure}
\begin{center}
\includegraphics[scale=0.4,angle=-90]{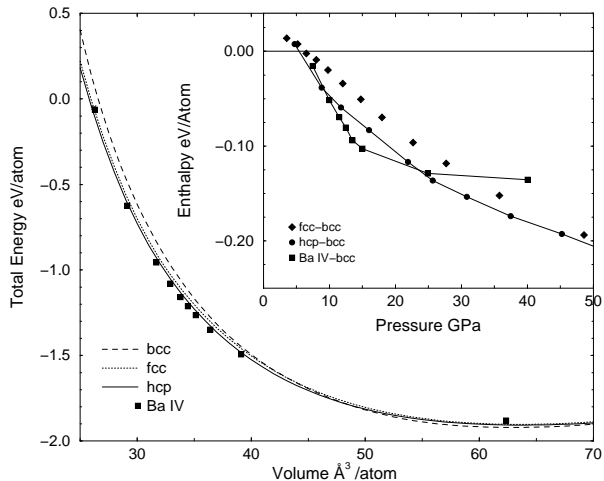}
\caption{\emph{Graph of energy against volume for various phases. At larger
volumes bcc has the lowest energy, while hcp is lower below
$\sim 48\pm 2$\AA$^3$. The inset shows the differences in
enthalpy between the hcp, BaIV and fcc with reference to that of bcc.
The calculated Ba IV data is shifted by -0.065eV (see text) to 
account for  the incommensurability.
}}
\label{fig:EngVol}
\end{center}
\end{figure}
\setlength{\columnwidth}{\oldcolwidth}
}

\newcommand{\registrygraph}{
\setlength{\oldcolwidth}{\columnwidth}
\setlength{\columnwidth}{\newwidth}
\begin{figure}
\begin{center}
\includegraphics[scale=0.35]{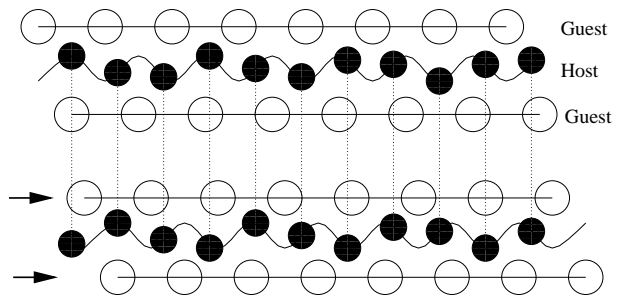}

\vspace*{0.05in}
\caption{\emph{In an incommensurate structure, guest atoms occur at all
possible heights in the host unit cell.   The guest-host interactions
could induce a modulation with period $c_{guest}$ in the host 
which would yield an 
indirect chain-chain coupling (upper figure, with the host displacements
much exaggerated).  
Since all local registries are observed
somewhere along the chain, the energy \emph{change} due to a
vibrational mode displacing the guest chains along their length with
respect to the host  is exactly zero (lower figure).  The host
modulation would move with the  guest atoms holding  adjacent chains
in registry with one another. 
}}
\label{fig:registry}
\end{center}
\end{figure}
\setlength{\columnwidth}{\oldcolwidth}
}



\newcommand{\ctoaratiograph}{
\setlength{\oldcolwidth}{\columnwidth}
\setlength{\columnwidth}{\newwidth}
\
\begin{figure}
\begin{center}
\includegraphics[scale=0.4]{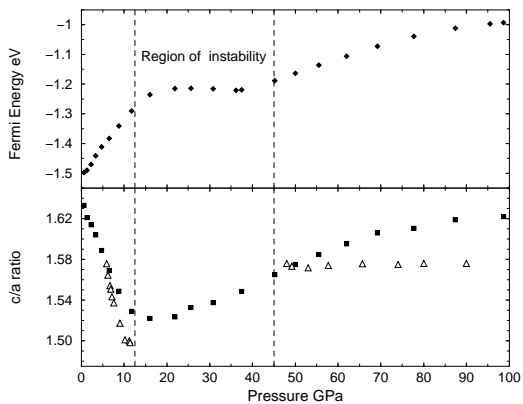}
\caption{\emph{
Pressure dependence of the Fermi energy (top) and c/a ratio (bottom) 
for hcp Ba, from constant (hydrostatic) pressure calculations. The
open triangles are experimental data taken by Takemura
\protect\cite{Takemura}. 
}}
\label{fig:ctoa-ratio}
\end{center}
\end{figure}
\addtolength{\leftmargin}{-1cm}
\setlength{\columnwidth}{\oldcolwidth}
}

\newcommand{\nearestgraph}{
\setlength{\oldcolwidth}{\columnwidth}
\setlength{\columnwidth}{\newwidth}
\begin{figure}
\begin{center}
\includegraphics[scale=0.4,angle=-90]{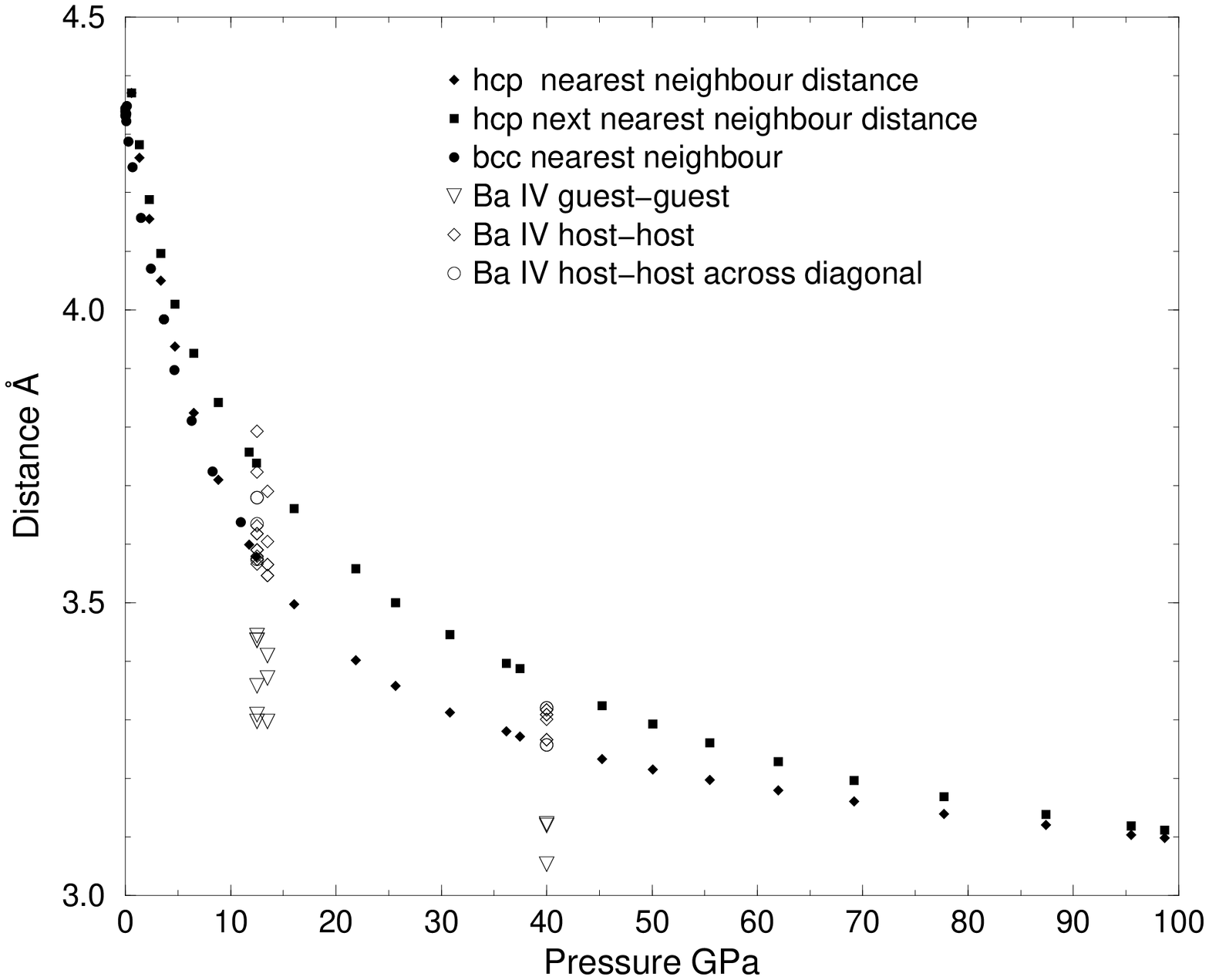}
\end{center}
\caption{\emph{Observed bond lengths as a function of pressure in hcp, bcc and Ba IV
structures. The 
Ba IV interhost distances are similar to the neighbor
distances in hcp, which themselves are a continuation of the bcc distances.
The Ba IV guest separation is consistently lower. The
shortest possible guest-host distance is half the distance across the host
octagon, which is slightly more than the smallest guest-guest spacing.
Hence the guest atom is more compressed than the host.}}
\label{fig:bond-lengths}
\end{figure}
\setlength{\columnwidth}{\oldcolwidth}
}

\title{Theoretical and computational study of high pressure structures in barium}
\author{Stewart K. Reed and Graeme J. Ackland}
\address{Department of Physics and Astronomy, The University of
Edinburgh\\
Edinburgh, EH9 3JZ, Scotland, United Kingdom}

\newcommand{\host}[1]{$#1_{host}$}
\newcommand{\guest}[1]{$#1_{guest}$}
\newcommand{\guestc}[1]{$#1_{guest_t}$}
\newcommand{\guestm}[1]{$#1_{guest_m}$}
\newcommand{\super}[1]{$#1_{s}$}
\maketitle
\makeatother

\begin{abstract}

Recent high pressure work has suggested that elemental barium forms a
high pressure self-hosting structure (Ba IV) involving  two `types' of 
barium atom. Uniquely among reported elemental structures it cannot be described 
by a single 
crystalline lattice, instead involving two interpenetrating 
incommensurate lattices.  In this letter we report pseudopotential
calculations demonstrating the stability and the potentially
disordered nature of the `guest' structure.  Using band structures and
nearly-free electron theory we relate the appearance of Ba IV to
an instability in the close-packed structure, demonstrate that it has
a zero energy vibrational mode, and speculate about the
structure's stability in other divalent elements.

\end{abstract}


\begin{multicols}{2}

The high pressure structures of the alkaline earth metals show  an 
unusual trend in going from apparently close packed
structures at low  pressure to more complex open structures at higher
pressure.  This counterintuitive behavior arises from transfer of
electrons from free-electron like  s-bands to more directional d-bands.

Barium has a number of still more unusual features. It adopts
the body-centered-cubic (bcc) structure at zero pressure
(conventionally denoted phase I).  Under  5.5 GPa of pressure it
transforms to phase II with a  hexagonal-close-packed structure (hcp).
At 12.6 GPa it transforms into the complex phase IV and then back to hcp
(phase V) at 45 GPa. This last phase is observed experimentally 
up to at least 90 GPa.  Until recently all attempts to solve the
phase IV structure have failed.

Nelmes 
{\it et al}\cite{NelmesAllan} have now shown that Ba IV has a complicated
structure consisting of two interpenetrating but incommensurate structures:
a tetragonal `host' with `guest' chains
in channels along the $c$-axis of the host. Although they share the same 
$a$ and $b$ parameters the guest and host lattices are incommensurate 
in the $c$-axis.
Viewed along $c$, the eight atoms in the host unit cell are arranged in
an octagon but with atoms at alternating heights. 
The guest chains pass through the middle of these
octagons, and a combination of diffuse scattering and Bragg peaks
suggests regions of both order and disorder between the chains.  
Figure \ref{fig:hostguest} illustrates the unit cells 
and describes how Ba IV  splits up in distinct `subphases' depending 
on the registry of adjacent chains.

An unusual feature of the hcp phase is the strong pressure dependence
shown by the c/a ratio in phase II: it falls to $\sim$1.50 at the II-IV 
transition, so far
below the ideal value (1.633) that the structure can hardly be regarded 
as close packed. In phase V however, which is also hcp, the reported c/a ratio
is almost pressure independent at $\sim$1.58 \cite{Takemura}.


In this paper we present total energy calculations on the observed
and other possible structures.  These enable us to probe
experimentally unobservable quantities such as the energy barrier to
moving the chains and
whether there is an isostructural
II-V phase transition `hidden' in the range of pressures
where Ba IV is more stable. Dealing with the 
incommensurate double-lattice provides a unique challenge for 
simulations based on periodic boundary conditions.  
\hostgueststructure

The calculations 
are done using the ab initio total-energy  density-functional-theory
\cite{HohenbergKohn,KolnSham,LS} plane-wave ultrasoft pseudopotential method
\cite{PayneEtAl,Vanderbilt} which is  well documented elsewhere. 
%
The generalized gradient approximation\cite{P:PerdewWang} was used for
exchange and correlation. The plane wave basis is cut off at
350 eV which is sufficient to converge total energies
to better than 0.1 meV per atom. Sampling of the reciprocal space
uses symmetry-reduced k-point sets\cite{MonkhorstPack} converged to 
0.1 meV per atom.
Eight valence electrons per atom are treated explicitly.
We calculate the total energies for the bcc, fcc, hcp and Ba IV
structures.  In all cases the enthalpy is minimized with respect to 
ionic and unit cell degrees of freedom under hydrostatic pressure
using Hellman-Feynman forces and analytic stresses with Pulay 
corrections\cite{mcw}.

Plane wave calculations are done with periodic boundary conditions,
but it is impossible to model an incommensurate structure exactly
using a supercell.  Hence Ba IV was approximated by a similar
commensurate structure: pseudo-Ba IV comprises a monoclinic supercell
(s) containing eight host and three guest atoms with
\mbox{\super{a}$=\frac{1}{2} ( b_{host} +
  a_{host})+\frac{1}{2}c_{host}$}, \mbox{\super{b}$=\frac{1}{2} (
  b_{host} - a_{host})+\frac{1}{2}c_{host}$} and
\mbox{\super{c}$=2c_{host}$}. Before relaxation three guest atoms were
evenly distributed along the $c$-axis from the origin. This
arrangement gives a \host{c} to \guest{c} ratio of 1.5 compared to
experimental value of about 1.39. Furthermore, since \super{a} and
\super{b} both contain a component in the \host{c} direction the
neighboring columns of guest atoms are displaced in the \host{c}
direction with respect to each other more than is seen
experimentally\cite{mono}.  The total energy for this cell will
represent an upper bound on the energy of incommensurate Ba
IV.
\energygraph
The results of the calculations are shown in figure
\ref{fig:EngVol}. The energy differences between the bcc, fcc and hcp
structures are small but they are nonetheless greater than the
accuracy of the calculations. In agreement with experiment, bcc 
is the favored structure at ambient pressure with  a 
phase transition to hcp at around 5 GPa.
The pseudo-Ba IV phase  energy is  close to the 
hcp curve,  moving away at low and high pressure: evidence 
of the reentrant nature of the hcp phase.

The overestimate of the energy in calculating pseudo-Ba IV rather than
incommensurate Ba IV can be approximated by a simple elastic model:
assume that to form pseudo-Ba IV the Ba IV guest lattice has been
independently compressed and the host lattice expanded until the
experimentally measured ratio
$c_{host}/c_{guest}=$1.39\cite{NelmesAllan} reaches the commensurate
value of the supercell: 1.5.

By measuring the stress $\sigma_{33}$ of the commensurate cell under
uniaxial compression $\epsilon_{33}$  and assuming that guest and
host lattices separately have the same elastic constant 
$C_{33}$, linear elasticity yields a strain energy of:

\vspace*{-0.05in}

\[  \Delta E = {\small \frac{V}{4}} C_{33} [1-2c_{host}/3c_{guest}]^2 \]

\vspace*{-0.05in}

The composite elastic modulus $C_{33}$ can be measured by a finite
strain\cite{bbk} calculation, taking a pressure independent $C_{33}$=150GPa 
gives the energy shift $\Delta E$ = 0.065eV/atom  which is 
included in figure 2.  The Ba IV enthalpy lies below hcp between 
9.5 GPa and 23 GPa, in reasonable agreement with experiment.  
The exact transition pressures are sensitive to this correction.
\registrygraph

The offset of the origin of the two lattices must be defined for
commensurate structures: we used a range of origin offsets and found
that the energies varied between them by less than 0.001eV/atom and
the relaxed volumes by less than 0.02\%\cite{offset}.  This shows that
no position of the guests is more favorable than any other which means
that the guests can be at any height with respect to the hosts without
there being an `energy penalty' to pay.  Thus the \guest{c} is
determined by the packing along the chain rather than host-guest
interactions.  In the fully incommensurate case this would lead to a
zero frequency phonon mode, or phason (figure \ref{fig:registry}).

The calculated c/a ratio was  0.58 as compared with experiment
\host{c}/\host{a} value of 0.56.  This can be explained by the elastic
model: 0.2 more guest atoms have been forced into the cell than would be ideal,
applying a stress on the host.  The host
cell therefore responds by expanding in the c-direction.

Moreover, hcp calculations show  a plateau in the Fermi energy
between 18 GPa and  43 GPa resulting from
a peak in the density of states.  Typically, this would lead to a 
Jahn-Teller type instability of hcp.

Recent studies \cite{Takemura,WinzenickHolzapfel} of the two hcp
phases have shown a dramatic decrease in the c/a ratio in phase II
with increasing pressure. In phase V it is noted that the ratio is
almost constant, slightly less than
ideal.   Experimentally it is impossible to get data between
12 GPa and 45 GPa because hcp is unstable in this
region; it is nevertheless possible to do calculations.  
\ctoaratiograph
As figure \ref{fig:ctoa-ratio} illustrates our c/a results initially agree
very well with experiments and remain in good agreement up to the phase 
transition.  In the experimentally inaccessible region we find a steady 
increase in c/a, and decrease in volume until once again 
good agreement is obtained when the
hcp phase is again observed at 45GPa.  Thus we have demonstrated that phase II
and phase V are actually the same phase, with the variation in c/a due
to a continuous transfer of electrons from $s$ to $d$.  
Zeng et al \cite{Zeng} calculated the c/a ratio using the linear 
muffin-tin orbital method
for electrons in a variety of different configurations. They found
that, at 5.7 GPa near the onset of phase II, a better fit to the
experiments was obtained if they included d-orbitals as well as s- and
p-orbitals.  They attribute the variable c/a ratio to strong 
s $\rightarrow$ d transfer across phase II, but report a much 
smaller effect in phase V. 

To interpret these results, we consider 
packing of hard spheres and  the nearly free electron model. 
\cite{hafner}.
As the host and guest cells are incommensurate the guest atoms must be
able to take take any (every) height with respect to the host cell.
In the packing model this implies that the guest
atoms must be able to pass through the
octagonal host rings. With hard sphere atoms 
this would mean either the host atoms cannot be in mutual
contact or the guest atoms must smaller than the hosts.  Even then,
the packing fraction is 0.56 whereas, at the II-IV transition the hcp
phase II has a packing fraction of 0.68.  This suggests that efficient 
packing is insufficient to understand the structure.

In the electron gas/pseudopotential perturbation theory 
model of Heine and Hafner \cite{hafner}, 
screening of the nuclear charges gives rise to an oscillatory
effective pair potential between atoms.  The 
favored crystal structure is determined by this
pair potential, and for Ba the high pressure close 
packed structures have atoms at unfavorable separations, near 
maxima in the pair potential\cite{Zeng}.  
Thus the `ideal' hcp structure is unstable with respect
to splitting the shell of 12 nearest neighbors into 6 and 6 by altering the
c/a ratio.  

\nearestgraph
Since the potential depends only on interatomic spacings, competing
structures will have neighbor separations similar to the optimally
distorted hcp.  In our calculations ( figure \ref{fig:bond-lengths})
we find that bcc and hcp have approximately the same nearest-neighbor
distance at room temperature.  As the pressure increases the hcp
nearest neighbor distance splits into two due to the non-ideal c/a
ratio.  At the II-IV transition the intra-guest distance is 3.41\AA,
and the shortest intra-host distance is 3.43\AA. These distances are
close to the nearest neighbor distances in hcp (3.46\AA).  As a result
of the incommensurability, a range of host-guest separations
nearest-neighbor distances should occur.  The similarity of
bondlengths (figure \ref{fig:bond-lengths}) to those of hcp shows that
Ba IV structure represents another way of arranging the atoms such
that they lie in a minima of the screened electrostatic potential.

An alternate view of the nearly free electron picture\cite{jones} is
that the energy is lowered by perturbation of states near
the Fermi surface: this favors structures which
have Brillouin zone faces close to the Fermi vector.
Incommensurate Ba IV has two Brillouin zones with the $(220)$, $(211)_{host}$
and $(111)_{guest}$ facets of the first zones close to the divalent 
free electron Fermi vector\cite{Degty}.

In sum, the total energy pseudopotential method gives an excellent
description of the experimentally observed high pressure phase diagram
for barium, including the complex guest-host structure Ba IV.
Although one can not represent the full incommensurate structure in a
supercell calculation, the energy of an equivalent commensurate
pseudo-Ba IV structure is close to that of hcp: treating the mismatch
using linear elasticity shows that incommensurate Ba IV has a region
of stability.

Calculations involving moving the guest chains within their channels,
show the interchain interactions of the guest atoms to be very weak,
probably mediated by elastic strain in the host.  Based on this we
propose that the tetragonal phase\cite{NelmesAllan} is metastable.
Weak interactions mediated by strain in the host lead to ordering at
low T.  The least favored arrangement in terms of strain would be to
have all guest atoms simultaneously coplanar, and so each chain is
displaced relative to its neighbors leading to the monoclinic
phase\cite{ortho}.
 
Ba IV an unusual example of a stable elemental structure with two
inequivalent atomic sites.  Pressurization of barium proceeds by
transfer of electrons from low energy, extended s-states to higher
energy, more localized $d$-states.  The hcp structure represents one
method of achieving intermediate $sd$ bonding: hybridized orbitals
with all atoms equivalent.  Ba IV offers another mechanism, with a
division between guest atoms containing more $d$-like character and
host atoms having more $s$-like behavior: a projection of the
Kohn-Sham wavefunctions onto atom centered orbitals shows $s-d$ transfer 
in all structures and a 20\%
greater $s$-character on the host atoms compared to the guest atoms
in the region of Ba IV stability.
The $p$ and $d$-character discrepancy was less pronounced, suggestive
of a transfer from $s$ to more free-electron like behavior.  The
larger number of electrons `localized' on the host atoms is consistent
with their apparently smaller size.  Thus Ba IV can be regarded as an
intermetallic compound, in which both components are actually the same
element.

The free electron theory suggests that Ba IV will be a competitor with hcp
as a high pressure phase not just in Ba, but in other materials with
similar valence electron density and  $s\rightarrow d$
electron transfer.  Moreover the reduced size of the $d$-like atom will
lead to faster diffusion and lower melting points at high 
pressures.

We would like to thank K.Kamenev, C.Verdozzi, D.R.Allan, R.J.Nelmes,
V.F.Degtyareva, M.I.McMahon, V.Heine and A.N.Jackson, for illuminating 
discussions,  EPSRC for support under grant GR/L68520 and a studentship (SKR).


\end{multicols}

\begin{thebibliography}{1}
\bibitem{LS}The calculations omit zero-point phonons and relativistic 
effects in the valence electrons, but the energy differences due to 
these are expected to be extremely small.  Calculations using 
spin-dependent functionals indicate that Ba IV is non-magnetic.
\bibitem{mono} The supercell corresponds to a displacement of adjacent
chains by $c_{host}/2$, which in turn corresponds to a monoclinic angle
of 98$^\circ$ compared with 96.15$^\circ$ reported experimentally.
\bibitem{offset}
The simulations in fig \ref{fig:EngVol} were carried out with 
host atoms at
$0$\host{c}, $\frac{1}{2}$\host{c}, \host{c} and
$\frac{3}{2}$\host{c} and guest atoms at $\frac{5}{12}$\host{c},
$\frac{13}{12}c_s$ and $\frac{21}{12}$\host{c}. 
Two additional calculations at 12.5GPa with the lowest guest atoms initially 
at $\frac{7}{60}$  and at 0  were done. The
volume of the  former was 0.02\% smaller than that found with the
guests starting at $\frac{7}{60}$\host{c} while the latter was 0.01\%
smaller.  These offsets must be specified to perform the calculations,
but represent a choice of position  along the incommensurate chain 
rather than a measurable quantity.
\bibitem{ortho} More complex structures involving alternate chains moving
up and down are also consistent with our results.
\end{thebibliography}
\end{document}